# PMMP – PQC Migration Management Process* - preprint -

Nils von Nethen, Alex Wiesmaier, Nouri Alnahawi, and Johanna Henrich

Darmstadt University of Applied Sciences, Germany

**Abstract.** Organizations have to plan on migrating to quantum-resilient cryptographic measures, also known as PQC. However, this is a difficult task, and to the best of our knowledge, there is no generalized approach to manage such a complex migration for cryptography used in IT systems that explicitly integrates into organizations' steering mechanisms and control systems. We present PMMP, a risk-based process for managing the migration of organizations from classic cryptography to PQC and establishing crypto-agility. Having completed the initial design phase, as well as a theoretical evaluation, we now intend to promote PMMP. Practitioners are encouraged to join the effort in order to enable a comprehensive practical evaluation and further development.

## 1 Introduction

Cryptographic mechanisms have always been subject to threats from advances in technology. Sufficiently large quantum computers (QC) pose such a threat, especially to the established asymmetric schemes that are rendered useless by quantum algorithms [33]. Therefore, cryptographers worldwide are developing quantum resilient cryptographic schemes, called post-quantum cryptography (PQC). The ongoing NIST PQC standardization process [10] aims at establishing standard PQC schemes to be integrated into IT systems. However, adapting and migrating large software infrastructures to PQC is a difficult task that is accompanied by several requirements and challenges.

We present the PQC-Migration-Management-Process (PMMP), a novel organizational and practical framework for managing the migration to PQC in organizations and IT systems. It provides a step-by-step guide towards achieving crypto-agility and successfully migrating towards PQC. PMMP also addresses non-migratable cases by allowing for suitable bridging approaches, depending on the organization's IT infrastructure or use cases. PMMP is thought to be integrated into an existing IT security management system (ISMS) such as ISO 27001 [11], and works best when done so.

PMMP requires defining a timeline for getting ready to handle the quantum threat, which should be individually determined based on an organization's risk assessment and specific security needs. PMMP relies on the existence of (standardized) PQ secure schemes, libraries, and protocols. The previously mentioned aspects, as well as the advancements in the development and standardization of PQC and its secure implementations, have already been thoroughly studied in previous works, e.g. [1–3]. These are not in the scope of this paper.

In the related work in Sec. 2, we present relevant PQC migration recommendations and exiting migration management approaches. Based on that, we identify the requirements and challenges for a successful migration in Sec. 3, as well as the scope and limitations of

* This research work has been partly funded by the German Federal Ministry of Education and Research and the Hessian State Ministry for Higher Education, Research and the Arts within their joint support of the National Research Center for Applied Cyber-Security ATHENE.

organizations. Building upon this, we present the design of PMMP in Sec. 4. We perform a theoretical evaluation of PMMP based on our previously defined criteria and also based on the requirements of the Crypto Agility Maturity Model (CAMM) [15] in Sec 5. Note that the work at hand presents the results of the initial theoretical design phase. In a next step, it has to be applied in actual organizational migrations to incorporate practical feedback into the process.

## 2 Related work

We present recommendations and overview papers that are rather general and may include technical advice that complements our work. In addition, we introduce existing migration approaches and frameworks that are similar or relevant to our work. A comparison of the related work to our work is given in Sec. 5.2.

NIST is one of the major institutions to address the migration planning towards PQC, considering it an essential aspect, which extends their focus beyond the mere standardization of PQC schemes. They provide a white paper [4] on the impact of QC and the related challenges, as well as a project description [5, 25] for the migration towards PQC.

ETSI's Migration Strategies and Recommendations to Quantum Safe Schemes [12] include a three-staged approach for the migration. These are: Inventory compilation, preparation, and execution. Each stage consists of multiple steps, as well as notes and recommendations for planning and execution.

The Public Key Infrastructure Consortium (PKI Consortium) has shown great interest in migrating towards PQC and held its first conference [30] on that topic in March 2023. They focus on the standardization efforts and recommendations made by organizations such as the European Telecommunications Standards Institute (ETSI) and the Internet Engineering Task Force (IETF), and PQC for PKIs and certificates. Additionally, they published a PQC Capabilities Matrix [29] (PQCCM) as a living document. It provides valuable information on libraries, hardware, and applications w.r.t. their PQC capabilities.

In a perspective paper [19], Joseph et al. offer an overview of the QC threat and the ongoing PQC development and standardization efforts. They also provide recommendations to organizations including the importance of crypto-agility, risk prioritization, and the use of hybrid solutions.

Another perspective paper [20] by Ma et al. discusses the need for crypto-agility and provide an outline of a framework to evaluate the risks resulting from the lack of crypto-agility. The proposed framework consists of five phases including identification, asset inventory compilation, risk estimation, risk mitigation, and roadmap creation. The authors provide a case study on the usage of their framework to asses and mitigate the QC threat within an organization.

Alnahawi et al. present a survey [3] on research on PQC so far. They describe challenges as well as already available solutions for PQC-enabled protocols. The website and its references were extensively used while writing the paper at hand. In another survey [1], they present an overview of the current state of crypto-agility. They shed light on different definitions, needs, and scopes of crypto-agility and evaluate the state of readiness.

The German Federal Office for Information Security (BSI) provides recommendations for migrating to PQC in [9]. They suggest enabling crypto-agility through hybrid soluions to react to changing security levels of the used cryptography, but how to achieve the needed levels of crypto-agility is not explained.

In [18], ISARA compares the migration to PQC to the migration needed for the Y2k-bug. The migration is driven by risk management, thus involving the leaders of organizations. Their approach distinguishes between the quantum riskand the cryptographic risk.

To manage the quantum risk, the organization needs to become crypto-agile, which is quite similar to managing the cryptographic risk as a whole, but instead focusing on the specific needs of PQC.

The PQC Migration Handbook [35] provides advice and concrete actions for organizations wishing to migrate to PQC. It is directed towards IT security practitioners and their management. It follows a three step migration approach (diagnosis, planning, executing), and underpins them with concrete action items.

In [40], Zhang et al. discuss the lessons learned from migrating an IBM Db2 database to using PQC. The approach is inspired from the migrations mitigating the Y2K-bug. They propose to get prepared by investing in crypto-agility and using software design patterns such as the factory pattern. Regarding the problem of coordinating with business partners, the authors refer to the already ongoing work on Internet standards featuring PQC [17] by IETF. They state that migrating to PQC is a community task that software practitioners have to handle, primarily by upgrading the software they maintain to be crypto-agile. One of the most demanding challenges they encountered while migrating the IBM Db2 database was the lack of documentation quality. Additionally, they often encountered hard-coded key lengths that needed to be updated.

In [21], Mashatan and Heinzmann advise establishing a governance model and body to follow their recommendations for migration. Following that, they advise assessing the risks QC might pose to the organization. Finally, concrete PQC schemes should be selected and implemented.

Hasan et al. [14] present a framework to aid organizations through the migration to PQC that can be integrated into enterprise settings. The framework consists of six steps. These are creating a crypto-inventory, creating a data inventory, classification, evaluation, recommendation, and review and maintenance. They also provide a security dependency analysis process, which comprises eight steps covering aspects such as communication, protocols and primitives, and (virtual) machines and containers.

White et al. [38] report of the IBM Z transition to quantum-safe technologies in an extensive manner. The self published book consists of a general guideline, in addition to a specific migration conducted on the IBM Z system. The general migration recommendations wrap up the experiences and lessons learned from their own executed migration focuing on education, crypto-inventory, and crypto-agility. It also sheds some light on organization business process aspects.

## 3 Scope, Limitations and Requirements

On the surface, the scope of the PQC migration may appear to be very similar to a regular adoption of any new technology or cryptographic mechanisms. However, the new breed of PQC primitives poses some serious limitations on both the organization's capabilities and their IT-infrastructures for the integration of these new schemes. Thus, the scope of the migration should involve not only knowledge about the what and where, but also the how. Limitations may include, for instance, larger key sizes unsupportable by the used hardware, or higher latency and packet loss rates that render an application non-usable in practice. Such aspects are thoroughly addressed and discussed in the literature provided in Sec. 2. Based on the existing works and our own contribution, the requirements of the developed migration management approach to PQC are described in the following:

*Migration timeline* As stated in [36], the integration of PQC can and should start today. In [18], it is recommended to start early, since cryptographic upgrades are challenging and time-consuming. Similarly, [37] suggests putting the topic on the agenda in order to prevent harvest-then-decrypt attacks Furthermore, [39] advises having a clear roadmap and a timeline for the migration. Consequently, PMMP needs to help define a timeline for the migration. Executives must be able to estimate the duration of each step and the migration as a whole. Therefore, each phase of the migration has to have metrics by which the duration can be measured, estimated, and steered.

*Security* The migration process must not lead to the emergence of new vulnerabilities. E.g., disabling cryptography modules completely, because they cannot get migrated, must not be an option and must be prevented by the process with appropriate countermeasures. Applications offering both classic cryptography and PQC must not allow downgrading from PQC to classic. Exceptions will be needed, but have to be reviewed and accepted. Moreover, the newly implemented PQC algorithms have to be used correctly. E.g., the process might include educating participants in the migration (e.g., programmers, administrators, or project leaders). Furthermore, in [36], it is recommended to use hybrid methods that use pre- and post-quantum cryptography simultaneously. Similarly, [18] proposes to favor hybrid methods for higher-risk applications and promotes crypto-agility in the organization. In [22], it is suggested to "raise hybrids", taking the best of both worlds and enabling interoperability between systems. Therefore, it can be assumed that hybrid solutions will probably become the de facto standard for integrating PQC.

*Completeness* The process must ensure that all relevant systems that need it get migrated to PQC. The migration process has to recommend a mechanism by which a comprehensive list of systems needing a migration can be compiled [39].

*Context awareness* In [18], it is recommended to ask vendors about the "quantum-readiness" of their products. For the migration process, this means that one needs mechanisms that assess the crypto-agility of vendors if third-party systems are used. Most importantly, the migration to PQC can only succeed if communication partners migrate as well. The migration process is required to ensure that a context-aware migration strategy is developed and applied. Additionally, in [40], the authors report the need to acquire substantial knowledge on the topic before developers can start the actual work. In [28] it is pointed out that employees must be educated in the new technology. The migration process has to elaborate on how to educate involved personnel. E.g., it is suggested to collaborate with universities as these provide a large pool of talents. This could be used for the education of the employees and for migration support. Approaching the PQC era, there will no longer be one universal algorithm per task that can be used for all use cases. the migration has to be tailored to the needed security level of the protected data [36]. The decision on which algorithm is used has to be balanced between speed, size, and security. Consequently, the migration process has to guide the selection of suitable algorithms for different applications.

*Crypto-agility* Technical debts, such as hard-coded key lengths, add further levels of complexity [39]. The BSI recommends implementing crypto-agility while new applications are developed or existing ones are upgraded. This enables the organization to update the cryptographic primitives more easily [8]. Other organizations, such as ISARA [18] or utimaco [37] also recommend this. Thus, the migration process must present mechanisms that can help ensure a sufficient level of crypto-agility. This also allows for exchanging (unbroken) algorithms with newer ones, e.g., for efficiency reasons. Furthermore, the BSI recommends hybrid solutions that combine classic cryptography with PQC solutions [8]. For high security areas, the BSI even requires hybrid solutions [9].

*Interoperability and availability* While migrating an application, connected applications need to stay able to communicate with the migrated application. Interoperability must be ensured for the organization as a whole and must not interrupt the operation of the organization's business processes. As an example, in [18], it is required to implement "forward/backward compatibility" such that systems can still operate during migration.

*Interim results* Migrating an organization to PQC can take quite a long time. It must be possible to deliver interim results that can be used before the entire organization is migrated. This gradual migration can be achieved, e.g., through backward compatibility of the migrated subsystems.

*React to advancements in cryptanalysis* The process must provide solutions that can be applied when implemented algorithms turn out to be insecure. This is very related to the before-mentioned crypto-agility requirement. To react adequately to advancements in cryptanalysis, the respective state-of-the-art has to be known before, during, and ever after the migration process.

## 4 Defining the Management Process

Because the migration to PQC is a major change requiring substantial resources, e.g., powerful hardware or external support, approval by the organization's decision makers is needed. ETSI recommends creating the role of a migration manager [12]. In organizations that have an IT security management system (ISMS) according to ISO 27001 [11] implemented, this might be the CISO, who is in charge of managing the organization's information security. Fig. 1 depicts the PMMP steps and their interdependencies. It is highly recommended executing the steps in the order presented by the sections to ensure a smooth flow.

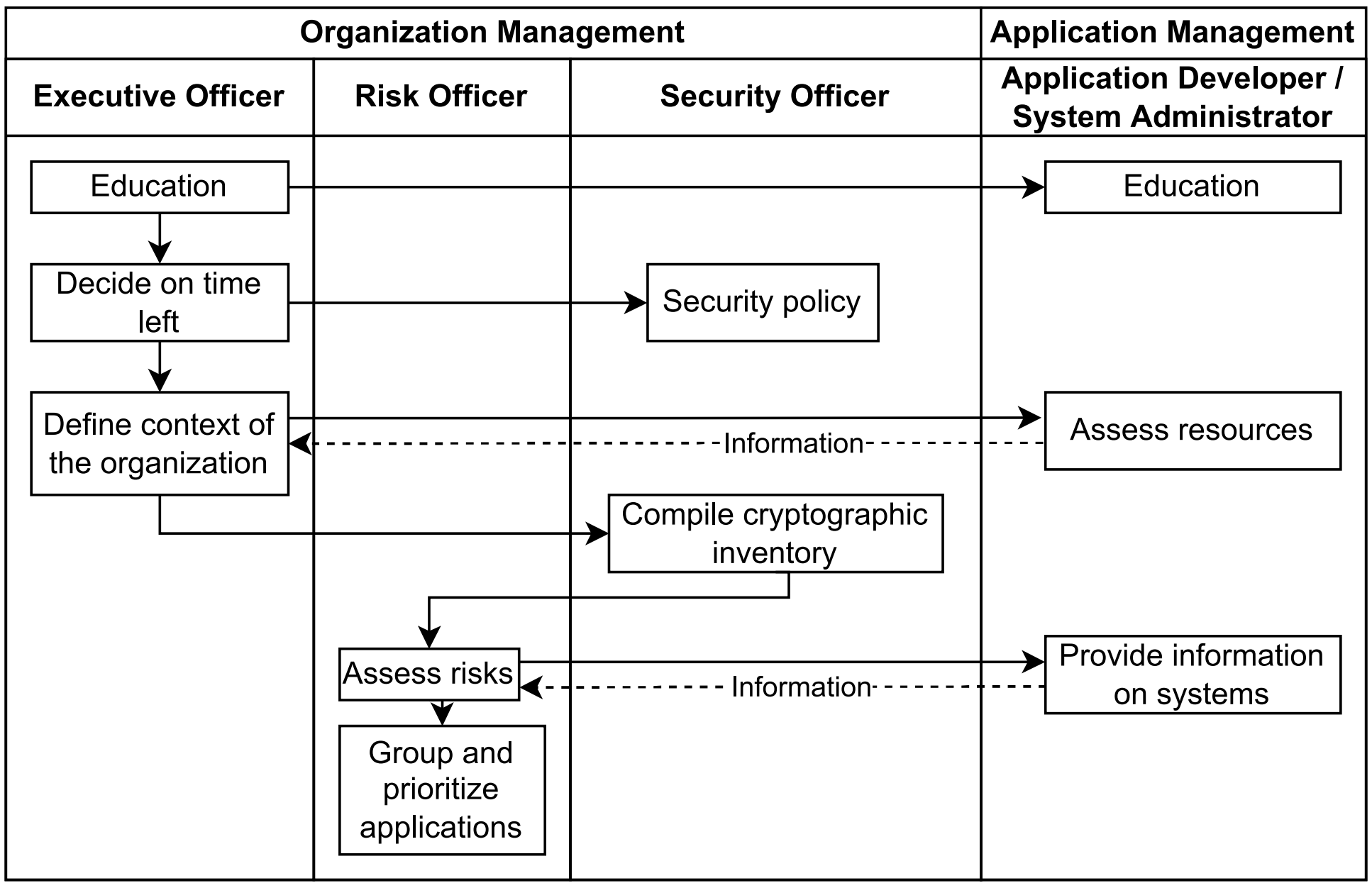

**Fig. 1.** PMMP Overview

### 4.1 Education

It is necessary to educate decision makers on the topic of PQC to achieve their agreement and support for the migration. Education can take the form of workshops or seminars where managers have the opportunity to ask questions and understand the dangers. Additionally, the software developers responsible for making changes to the organizationally hosted applications have to be educated. They must know how to implement the new algorithms or correctly use PQC libraries. If the knowledge of developing secure software is not yet established, engineers need to acquire it. Management has to provide funding to enable this type of security training (also [13]). The extend of the knowledge and/or skills needed for the migration differs between the roles (e.g., manager vs. developer) and actual tasks to be executed (e.g., ordering hardware vs. adapting software). In the end, the respective department heads are responsible for this decision. If the organization does not have adequate resources for providing that kind of education, it has to acquire external support.

### 4.2 Decide on time left

Following Mosca's theorem [24], the time left to migrate to PQC is limited by the time needed to develop, prepare and deploy the new schemes plus the time needed for secrets to remain secure. This allows the development of a timeline on which further decisions and the risk assessment are based. The decision has to be realistic and oriented at official estimates by authorities, e.g., ETSI [12] or NIST [4]. Nevertheless, it is important to point out the significance of the *store now - decrypt later* attacks, which may target currently non-PQC protected data, as well as future planned IT-projects [19]. Considering the sensitivity of organizations' data, a decision on the time left for migration should be made with such aspects in mind. That being said, there are no concrete estimates regarding when large scale QCs could appear (prevailing assumption by experts is around 20 to 30 years i.e., between 2040 and 2050 [23]). Based on these assumptions, and the factors mentioned earlier, the PQC Migration Handbook [35] provides a formula for making that decision based on the time organizations' assets need to remain secure, the migration time, and the time left until large scale QCs emerge (cf. Mosca's inequality [23]). Depending on the assumed urgency of a migration, this method offers a solid base for the migration timeline.

### 4.3 Security policy and goals

The PQC migration goals should be documented in the organization's goals and need to be published inside the organization. Additionally, the security requirements for reaching the goals have to be documented in an organization-wide security policy. Organizations applying an ISMS already have a security policy that should be extended accordingly. New software, either developed in-house or provided by third parties, can be required to feature a certain level of crypto-agility which can be used to (later) implement PQC. To set a migration timeline, the policy has to define the date on which CRQCs are thought to be a real thing. As shown in Fig. 1, the security officer is in charge of developing the security policy based on the management decisions. The policy can feature concrete technical statements with which algorithms the organization wants to protect which data. In addition, the security policy has to document a list of excluded algorithms if needed. This has to be coordinated with the communication partners of the organization. Alternatively, the security policy can reference a technical document from authorities [7].

### 4.4 Context of the organization

The context of the organization defines the scope in which migration tasks have to be applied. Below sections point out how the context of the organization can be defined and what it is made of. In Fig. 1, this step is situated in the column of the executive officer, who is in charge of defining the context of the organization.

*Stakeholders* As the success of an organization often strongly depends on data security, the stakeholders have an interest in information security. In effect, the organization can get obliged by its customers to migrate to PQC. This might be the case if a Service Level Agreement (SLA) exists between the organization and its customers, defining the level of security, availability, or similar parameters defining the service. The other way around: If the organization's service providers process data relevant to the stakeholders, it must be examined to what extent the service providers can be required by the organization to maintain the same level of security. Managing the migration to PQC in the organization would in this case also affect the suppliers of the organization, that can get obliged to migrate to PQC as well. Authorities can require organizations to fulfill legal or regulatory requirements, as is the case for financial institutes. Protection of customer data is required by data protection laws such as the European General Data Protection Regulation (GDPR) [34] that requires organizations to use state of the art IT security techniques. Some day, state-of-the-art technology will include PQC and require organizations to adapt their cryptography. As the stakeholders also have an economic interest in the success of the organization, they are required to agree upon the organization's plans for migration [28]. This is because the migration requires financial resources, e.g, to provide upgraded hardware or to get external support. Stakeholders can also be insurers who cover the organization against cyber risks. If standardized PQC algorithms are not used, the insurance may be invalidated. This might additionally be a driver of the migration.

*Communication partners* When migrating to PQC, communication partners have to commit to a list of used algorithms, because they need to stay compatible. Therefore, it is advised to not only compile a crypto-inventory as explained in Sec. 4.5, but also to compile an inventory of communication partners. The list of partners has to document what their communication endpoints feature, e.g., which cryptographic primitive(s) they are able to run, in order to exchange information with the organization. Ideally, the algorithms are supported on the hardware the partners are already using while fitting their needs. For proper identification of the algorithms, existing and standardized algorithm identifiers have to be used. In the same manner, the partners should exclude algorithms they do not want to use (because of trust issues and alike).

There are communication partners with whom the organization cannot agree individually on algorithms, e.g., users of a website. Depending on the need for protection, these connections must be encrypted opportunistically. For this purpose, the usual users of the website must be analyzed in order to determine which algorithms come into question. Then, they can be provided with the best possible security. On a technical layer, opportunistic security is implemented within the applications that are migrated. With a so-called friends-and-family phase, the migration of selected applications can be tested and evaluated. In this phase, only a few selected communication partners upgrade their systems to PQC. A successful prototypical migration can then be used to advertise the use of PQC and get other communication partners to migrate also.

*Resources* Management has to provide resources for the migration, e.g., investments in the education of the employees, external support, etc. Before money is spent, the organization should assess what it can start without acquiring more resources. That is, estimating which computing resources are used to which extent and how much room there is for algorithms with higher resource demand. Likewise, the resources of the software engineers need to be calculated. Are they able to implement new algorithms or securely use updates of cryptographic libraries featuring new cryptographic primitives? In Fig. 1 this relationship is shown by the dotted arrow connecting "Assess resources" in the rightmost lane and the "Define context of the organization" in the leftmost lane.

### 4.5 Inventory of cryptography

To be able to assess the risk and the impact of QC on the organization, the organization needs to compile a crypto-inventory, which serves as a starting point for the migration. All relevant systems must be discovered to ensure that no systems are left during migration. The inventory provides a list of applications linked to the type of cryptography, the algorithm and the key length used in each case. In Fig. 1, the creation of the crypto-inventory follows after defining the context of the organization. This is the task of security officers, since they have the resources to interpret the technical details of the inventory and document the security level of the applications.

The inventory can be compiled using various methods. E.g., if there are only a few systems, the inventory can be compiled by hand. For bigger environments, Alnahawi et al. propose the development and usage of automated cryptography detection tools [3], which is also recommended by NIST in [4]. These tools could scan systems, e.g., for stored SSH keys, libraries used in applications, or trusted root certificates of a PKI. Applications developed in-house by the organization where access to the source code is possible can be scanned using text-based tools. E.g., it might be sufficient to scan the source code for the terms `encrypt` or `decrypt` to find the used cryptographic primitives. It might also be possible to scan the whole network traffic in an organization to detect cryptography used in the network, e.g., TLS handshakes. Then the chosen algorithms and key lengths could be recorded [27]. Furthermore, documents about former security audits and interface descriptions of the systems can help compile a list of relevant systems and used cryptographic components. This sort of list can be organized into a pre-defined structure, such as the Cryptography Bill of Materials (CBOM) [16], which was recently proposed by IBM as an object model to describe cryptographic assets based on the CycloneDX standard for Software Bill of Materials (SBOM).

PKIs need special attention, as there are dependencies between keys [12]. A PKI can have many different entities and can be rather complex. All entities in the PKI have to migrate for the structure to function, as the certificates issued might be installed on numerous different devices. However, in a way, a PKI is already some kind of crypto-inventory itself.

After a group of applications (Sec. 4.7) was migrated to PQC, the crypto-inventory is updated. Then, the risk assessment can start again with the most recent state of the inventory. While compiling the crypto-inventory, it is important to document which data is protected by the used algorithms. To understand the application's data flow, reverse engineering of the system might be required.

### 4.6 Assess risks

PMMP aims to be risk-driven, which is why risk assessment is one of the most important steps for efficient and effective migrating. In Fig. 1, this step is situated in the lane of the risk officer, which is responsible for the risk assessment. The application developers (and maintainers) are responsible for providing information about the systems that are subject to the risk assessment. Therefore, there is an arrow connecting the two roles.

*Supporting documents* For each business process, the risks CRQCs pose need to be assessed. A business impact analysis can be helpful at this point, as it shows how the business is affected in case of interruptions to the services provided by the IT infrastructure of the organization. Moreover, it helps to justify expenses for the migration. Additionally, existing documents on business continuation can help to assess the risks. With the additional help of a crypto-inventory, the applications at risk can be identified, since the inventory shows which applications are using algorithms endangered by QCs. Assessing the risk involves evaluating how long the data protected by the algorithms used in the application needs to stay secure (cf. Sec. 4.2).

*Scope* While applying the aforementioned techniques, that is, using the results of the business impact analysis and the crypto-inventory, the risk assessment has to cover the whole scope of the organization. This includes the stakeholders of the organization, which might be business partners, the organization itself, or authorities. Not migrating to PQC can have consequences that stakeholders might or might not want to accept. E.g., a customer that feels their data is no longer adequately protected could terminate the contract with the organization or file a lawsuit. Authorities might impose severe penalties on the organization. Furthermore, communication partners that do not (want to) migrate to using PQC, might risk getting excluded from the business.

*Possible risks* An attacker with access to a QC might deprive the organization of its business. It is not possible to migrate all systems at once. For this reason, in some cases the risks must be accepted, at least temporarily, before migrating. Also, some algorithms providing a high level of security can slow down the initial connection to, e.g., a website for up to many seconds. The organization either has to accept the risk of losing customers because their website appears to be very slow, or upgrade the hardware, at least on their side of the connection. Also, it can turn out that replacing an application is easier. The post-quantum algorithms may turn out to be insecure during the migration. A required refurbishment or reverse-engineering of an application might prove to be too costly.

The organization's risk management provides the decision of which systems to migrate and when. Note that this decision is completely risk-driven. E.g., systems that operate in the inner network of the organization might not get migrated first, as the risk is not high. Although they could be migrated with a simple upgrade in less than a few minutes, they handle unimportant data and might be not reachable over the internet.

### 4.7 Risk- and process-based grouping and prioritization

Organizations use different applications to operate their business. The interoperability of the systems must be ensured to avoid the interruption of the organization's processes. The applications and systems are grouped by the business process they are tied to and are migrated one by one. In Fig. 1, the grouping of the applications is placed in the lane of the risk officer.

To recall the mitigation of the Y2K-Bug: It is important to remember that the organization only has limited resources of software engineers, which is why Putnam and Schultz propose to triage the systems needing a migration [32]. First, the systems that involve life and death need to be touched. Second, "if you are not in a life-and-death business" [31], the systems critical for running the organization's business need to be worked on. The systems to be handled last are those whose failure would be irritating, but not costly. Thus, the decision on which group is migrated first is based on the risk assessment done before.

### 4.8 Testing and monitoring

The migration to PQC is a risk-driven process that is integrated into the ISMS of an organization. Software migrations can take quite a long time. Therefore, the processes defined in PMMP need to be monitored for effectiveness. Being a cross layer aspect, the testing and monitoring step is not displayed in Fig. 1 for better readability. E.g., regulated organizations like financial institutes that already have an internal control system in place can use it to monitor the effectiveness of the developed migration process. E.g., a monitoring process might check if the migration processes defined can be applied correctly. The monitoring has to detect a lack of resources required for migrating, e.g., lacking knowledge in the field of PQC. In such a case, the management has to improve resources by organizing workshops, establishing cooperation with universities, or releasing more financial resources for the education of the developers. To ensure that the methods and techniques of PQC are applied correctly, security audits performed by external organizations can help. Also, internal security audits performed by a security officer are helpful.

## 5 Evaluation

We verify that PMMP meets the defined requirements, and compare it to the migration approaches presented in the literature (Sec. 2). This section is explicitly not about a practical evaluation of PMMP, but deals with the validation of the proposed theoretical concepts.

### 5.1 Meeting the requirements

*Migration Timeline* PMMP starts with educating senior management, to ensure the topic is understood in the organization. Then, the executives need to commit to a date by which CRQCs will be available and threaten the currently used cryptography. Based on the risk assessment and with the help of the crypto-inventory, a migration timeline can be defined. To allow estimating the duration of the migration steps, PMMP helps to approximate the amount of work by assessing the resources an organization has. Which application is getting migrated first is a risk-based decision. Using techniques like reverse engineering to understand an application's architecture, the amount of work needed for the migration can be assessed, e.g., by using the number of lines of code as an indicator for complexity. Also, the migration process allows falling back to replace legacy applications with new ones, if this is faster (or easier). To develop a migration timeline, PMMP requires setting the relevant dates stemming from Mosca's theorem [24]: The time when a CRQC will be available, the duration in which data has to stay secure, and the duration of the migration to PQC.

*Security* PMMP features the education of not only the decision makers but also of, e.g., developers that implement the new primitives. This provides a solid ground on which the algorithms can be applied correctly. PMMP tackles advances in cryptographic analysis by integrating into the organization's risk management. As soon as a cryptographic primitive is weakened, the situation is evaluated, since risk management and risk assessment are frequent processes.

*Completeness* PMMP makes use of the features of an established ISMS, such as system and structure analysis. This is complemented by the creation of a crypto-inventory that serves, among other purposes, the compilation of a list of migration-relevant systems. Note that the former and the latter tackle a common task from different angles, thereby complementing each other.

*Context awareness* By integrating into an ISMS, PMMP aims to be context-aware by involving business partners and customers early, ensuring the relevant communication partners have the same vision. Additionally, PMMP takes into account the needs of stakeholders and those of regulatory authorities. The selection of suitable algorithms for different applications is enabled by complying with respective recommendations, such as published by NIST or BSI.

*Crypto-agility* PMMP enables the establishment of crypto-agility by supporting the fulfillment of requirements for practiced crypto-agility as defined in the cryptographic agility maturity model (CAMM) [15]. A more detailed presentation of the respective CAMM requirements and how they are supported by PMMP is given in Appendix A.

*Interoperability* To prevent interruption of business processes, interoperability between the systems is important. PMMP advises the use of gateways between existing systems. Thus, systems that have not yet been upgraded to PQC can be connected using the gateway technology. Additionally, the migration process is based on a process- and risk-based grouping of the applications and systems. Thereby, the interoperability of the systems can be ensured.

*Interim results* The first step of the migration is to educate the executive management on the topic of QC and its risk to classic cryptography. The knowledge gained in this step is an interim result. Also, by raising awareness for the topic of PQC, a snowball effect can be triggered. When stakeholders are required to migrate to PQC, this triggers stakeholder's stakeholders to migrate also. This is an intermediate result in the global migration to PQC. Interim results affecting the organization can be reached by grouping the application according to their risk. Then, each group of migrated applications is an intermediate goal.

*React to advancements in cryptanalysis* The risk assessment of the organization is not a one-shot process. The risk is assessed frequently. Therefore, when advancements in cryptanalysis threaten PQC or the known classic cryptographic primitives, the migration process has to be applied again. When advancements are made during migration, it is adapted accordingly.

## 5.2 Comparison to existing migration approaches

The suggested migration plan by the NIST [5, 25] involves different scenarios, including libraries, hardware, and protocols. It also characterizes the desired architecture and security properties. However, this planning involves mainly "initial discovery steps for the development of migration roadmaps". In contrast to our work, NIST provides a broader view, such as the discovery of asymmetric cryptography in systems and interacting with standards-developing organizations to raise awareness of necessary changes.

A rather high level guideline for the PQC migration is presented in [5]. The project draft published by NIST describes five scenarios. These are: 1. FIPS-140 validated hardware and software modules; 2. Cryptographic libraries; 3. Cryptographic applications and cryptographic support applications; 4. Embedded code in computing platforms; 5. Communication protocols. All of which utilize cryptographic components vulnerable to the quantum threat. They also provide a high-level architecture as a list of components that should be considered in the risk assessment, in addition to the desired security properties for each. However, the project is restricted to the first discovery phase, and doesn't involve further steps for realizing the migration, as provided by the work at hand.

Similar to the approach presented in the paper at hand, ETSI's migration plan [12] requires the compilation of a cryptographic inventory and conducting a risk assessment before starting the migration process. It also addresses aspects related sub-system isolation, business processes and management. In contrast to PMMP, ETSI's approach considers trust and key management issues, in addition to other rather technical aspects.

The migration approach presented in the PQC Migration Handbook [35] also overlaps with this work. This can be seen in their focus on the first diagnosis phase, risk factors and migration urgency scenarios, migration execution, and the migration timeline as a whole. However, they do not provide a similar level of attention to details for the business processes and technical planning within an organization. PMMP has a stronger focus on the management aspects including the technical resources and communication partners of an organization. PMMP also aims at supporting the integration into existing ISMS and crypto-agility maturity models. On a different note, PMMP does not provide concrete steps for executing the migration of cryptographic primitives and protocols (e.g. public key encryption, TLS, and X.509 certificates). Such aspects are addressed extensively in the PQC migration handbook, and offer a valuable guide on a very technical level. These aspects were intentionally left out in PMMP, and we refer to general PQC-related recommendations and overviews (Sec. 2) which cover that in great detail.

One of the most mature approaches existing is the one developed by Zhang et al., presented in [39]. What their approach and PMMP have in common, is that the migration to PQC is triggered from the top of the organization, initiated by the decision makers that get educated on the topic. The second step is to compile a crypto-inventory, but it is not mentioned how the inventory can be compiled. Regarding the migration of one single application (the IBM Db2 database), the compilation of the inventory is not in their focus. Their step six requires the user of the approach to execute the cybersecurity policy. The policy controls the selection of appropriate solutions based on the requirements of the organization and budgets. However, it remains unclear how the requirements can be determined. While the approach by Zhang et al. suggests working on systems that handle critical data first, it is not explained, what critical data is. Here, a risk-based process would be better suited, as proposed in PMMP.

The migration approach presented in [21], similar to [20], features just as PMMP a risk-based approach. They present three different paths for migration that all lead to remediation projects. Organizations following path A will then observe their position in the field. Organizations following path B will then execute their roadmap, and organizations following path C will then "simply need to make (relatively minor) adjustments to what they already have in place". Additionally, the authors suggest the organizations also need to have a deprecation path for the pre-quantum cryptographic implementations in place. When the quantum risk for an organization is high, the approach suggests implementing hybrid cryptography as soon as possible. If the quantum risk is low, the approach proposes to wait for an update for the application(s) in question. The problem with this approach is that the risk assessment is done before compiling the crypto-inventory. It is unclear how the risk can be assessed if it is unknown which applications use which type of cryptography, which key length and for which data. PMMP puts risk management in focus and takes the advances in QC as a risk that needs to be mitigated by the organization. The basis for this process is the compilation of a crypto-inventory before assessing the risk, since the inventory is needed for the risk assessment. A common aspect in both works is involving the decision makers in all the phases of the migration planning and execution.

Similar to our work, the framework presented in [14] requires an exhaustive crypto-inventory, which can be compiled both manually and automatically. However, they add a data inventory and classification process, which are not necessarily required in PMMP. The following steps proposed in [14] focus on the technical details of dependencies and cryptographic components in a given system. The authors additionally provide three concrete case studies to evaluate the feasibility and suitability of their framework, most of which mainly address crypto-inventories and dependencies. Clearly, PMMP would greatly benefit from a practical evaluation based on a case study or the migration execution within an existing system.

The PQC migration of IBM Z [38] is by far the most concrete case study, providing both general recommendations, and practical measures for the migration execution. Similar to PMMP, the work emphasizes the importance of the senior level management and overall personnel education. They also deem a crypto-inventory accompanied with a vulnerability (risk) assessment vital steps in the migration process. In contrast to PMMP, they provide hands on lessons from their own migration experience, which can be seen as a double-edged sword. On the one hand, they present very valuable insights as to how to compile a crypto-inventory, identify risk factors, and how to gradually remedy them. On the other hand, their concrete measures are highly dependent on the IBM Z system, which might not always be applicable to other systems.

A common difference to all the above is that PMMP performs integration within existing management processes, such as a mature ISMS. It uses the steering mechanisms in an organization for the migration to PQC from top to bottom. While applying the steps of PMMP with the testing and monitoring in place, features of the internal control system are used. In that sense, PMMP complements the other approaches by providing a framework for integrating the various mirgation aspects into existing management processes.

## 6 Discussion

To get the migration on the way, it is needed to repeat the very first step of the migration process presented in this paper: Educate executives and decision makers in PQC. Only if the possible dangers to the cryptography used today are recognized, someone will take the money and change something. PMMP includes interim states and puts a strong emphasis on the context of the organization including its stakeholders and communication partners, with the latter being one of the most important aspects of a successful migration.

PMMP emphasizes the understanding of organizations of their cryptography: To be able to migrate to PQC, the organizations have to understand which applications are vulnerable and what level of security they need to provide for the data processed. I.e. they need to compile a crypto-inventory. To the best of our knowledge, to this day, there is no concrete tool that can be recommended for that, as the research is still ongoing. But, it seems likely that NIST will provide or procure one in the future [26].

Although with PMMP a migration process is now available, not every application can be migrated. Some applications cannot be migrated, e.g., due to license reasons or because the source code is not available. This problem can be solved using a technique presented by Sneed et al. [13], i.e. using gateways. E.g., a PQC-VPN can ensure that applications can communicate using PQC.

Organization have to assess resources they have to migrate to PQC. There exist resources that can easily be quantified, such as the financial backing or the computational power. But developer resources must also be assessed to estimate the need for external support. If there is a metric that can be used to assess the competence of software engineers, it should be used.

Recalling the similarity of the migration from IPv4 to IPv6: there are networks that have already switched to using the newer protocol, those that use both variants, and those that still only use the older variant. This development will very likely also be seen in the migration to PQC. In the beginning, there will be only a few networks migrated to use post-quantum key exchanges or encryption. In the meantime, there will be networks supporting both (or all three) possible variants: classic cryptography, hybrid cryptography, and PQC only.

Also, NIST will most likely publish the standards specifying PQC in the next two years (from 2023). With these standards, organizations can decide more easily which algorithms they need to implement. Further work on the topic might include using the developed migration process, improving it, and helping organizations migrate to PQC by increasing their crypto-agility. Meanwhile, we shall see how the global migration to PQC will move forward.

## 7 Conclusion

In the previous sections, PMMP, a process for managing the migration to PQC, was presented. To develop a process that is oriented to the needs of the industry, respective requirements were defined. The process was evaluated against these requirements, as well as compared to other approaches. The successful evaluation of PMMP suggests that it can be used for migrating from classic cryptography to PQC. It is explained which challenges have to be solved and how existing management methods can be used in the context of the migration. To achieve that, PMMP uses a risk-based method, regarding the advances in QC as a risk that needs to be handled by risk management.

In analogy to the development life cycle of maturity models presented by Becker et al. [6], the development of a migration process is an iterative process that includes feedback from outside experts and practitioners. PMMP needs to incorporate feedback from practice. Ideally, by applying it in real world cryptographic migrations and using the learnings to adapt the process accordingly. We encourage practitioners to join our effort and conduct with us this important phase of the process development.

## A CAMM requirements

CAMM [15] requirements are intended to measure (not establish) the crypto-agility of systems. We argue that PMMP delivers processes whose results fulfill all CAMM requirements up to level 3 (practiced).

### A.1 CAMM level 1: "possible"

*R1.0: System knowledge* PMMP fulfills this requirement by involving stakeholders of the organization and compiling a list of communication partners. By that, the organization understands its context and is able to evaluate the impact QCs might have on it. PMMP features a strong focus on risk management.

*R1.1: Updateability* PMMP fulfills this requirement by providing processes to supply the needed resources for necessary updates. E.g., the executive management is educated in cryptographic and quantum risk. Then, the management can provide monetary resources. To prevent restricting the functionality of systems, their requirements need to be clear. Then, it can be measured if any functionality is lost. PMMP involves reverse engineering processes where needed.

*R1.2: Extensibility* This requirement, in contrast to the requirement Updateability above, is not about assessing the ability to run/implement PQC, but actually about acquiring resources needed for the upgrade. Resources have to be approved to, e.g., buy new hardware, and allow for external support while migrating cryptography.

*R1.3: Reversibility* As the migration to PQC is managed per application, with each application getting migrated in a dedicated project, this requirement is met. Also, PMMP utilizes pilot systems to ensure the updates work as expected.

*R1.4: Cryptography inventory* PMMP defines processes for this requirement, that is how a crypto-inventory can be compiled in an organization. In conjunction with risk management, the level of security the cryptographic primitives provide and which level of security the data handled by the applications where the primitives are used is understood. Plus, the process compiles an inventory of communication partners.

## A.2 CAMM level 2: "prepared"

*R2.0: Cryptographic modularity* PMMP ensures that applications and systems of an organization are upgraded in groups based on their business process.

*R2.1: Algorithm IDs / R2.2: Algorithm intersection / R2.3: Algorithm exclusion* PMMP has a strong focus on the context of the organization, including analyzing its stakeholders and communication partners. In consultation with the communication partners and especially their capabilities in the changeover to PQC, PMMP enables a regulated changeover of the algorithms. The use of standardized algorithm identifiers is suggested for mentioned consultation. In addition, the intersection and exclusion of algorithms are determined in the reconciliations with the communication partners. Furthermore, risk-based regulations can ensure that certain algorithms are excluded.

*R2.4: Opportunistic security* PMMP also deals with this issue from the background of the communication partners. If no individual coordination can be made with numerous partners, as is the case, e.g., with the visitors to a globally accessible website, the process relies on opportunistic security. For this purpose, the users of the systems are analyzed beforehand and the best possible procedures are ensured.

*R2.5: Usability* With a risk-based grouping of applications and processes, PMMP ensures that an interoperability of the systems and thus the usability of all applications in scope of the migration is ensured.

## A.3 CAMM level 3: "practiced"

*R3.0: Policies* PMMP has a great focus on security management. E.g., the migration process initiates changes to an existing security policy. A policy is required to state what the organization wants in terms of migration to PQC. This clarifies which requirements applications need to fulfill. Policies are developed in reconciliation with the communication partners, that might be required to migrate.

*R3.1: Performance awareness* PMMP considers security and economic risks, such as losing customers who find themselves using a device performing a slow quantum secure TLS handshake. PMMP defines the process of either accepting the risks or stocking up hardware, at least on one side of the connection.

*R3.2: Hardware modularity* PMMP fulfills this application-specific requirement by including a refurbishment process in the migration process. If applications need to be migrated according to the risk assessment and the organization can influence SW and HW, the migration process ensures the required changes are made upon migration.

*R3.3: Testing* PMMP integrates into existing management processes like risk management. Since risk assessment is an iterative process, it is frequently checked whether the security needs of the organization are fulfilled.

*R3.4: Enforceability* PMMP is driven by the leaders of an organization, that is, the migration process is applied from top to bottom. A security policy put in place by the management ensures that the needed techniques, especially resources, are made available. Additionally, if resources are lacking, e.g., educational resources, the migration process recommends cooperating with universities and research institutions in the field.

*R3.5: Security* While PMMP is applied, the security of the organization is assessed using external security audits. To ensure the organization is not vulnerable to attacks, risk management assesses the security of the organization.

*R3.6: Backwards compatibility* PMMP focuses on interoperability, which includes being backward compatible with older systems. Additionally, PMMP presents different management strategies, e.g., the incremental packet conversion approach [13]. This ensures that the different parts of the application migrated stay compatible with each other. Moreover, the migration process makes use of various transition mechanisms.

*R3.7: Transition mechanisms* One technique that PMMP applies is the use of gateways. This idea is taken from the REMIP presented in [13]. Therefore, PMMP features processes to fulfill this requirement.

*R3.8: Effectiveness* PMMP uses project management techniques to ensure the effective upgrading of cryptographic primitives. Moreover, PMMP can be monitored by the internal control system. Then, e.g., lack of knowledge can be measured and reacted upon. Additionally, PMMP features an evaluation at the end of every migrated application. Also, PMMP provides solutions for systems that cannot get upgraded in time (and have to be replaced with a newer application).